\documentclass[conference]{IEEEtran}
\IEEEoverridecommandlockouts
\usepackage{orcidlink}
\usepackage{cite}
\usepackage{amsmath,amssymb,amsfonts}
\usepackage{algorithmic}
\usepackage{graphicx}
\usepackage{textcomp}
\usepackage{xcolor}
\usepackage[numbers]{natbib}
\usepackage{amsmath}
\usepackage{tabularx}
\def\BibTeX{{\rm B\kern-.05em{\sc i\kern-.025em b}\kern-.08em
    T\kern-.1667em\lower.7ex\hbox{E}\kern-.125emX}}
\begin{document}

\title{MD-Dose: A diffusion model based on the Mamba for radiation dose prediction
\thanks{Identify applicable funding agency here. If none, delete this.}
}
\author{
    \IEEEauthorblockN{
        Linjie Fu\orcidlink{0000-0002-9109-2670}\IEEEauthorrefmark{1}\IEEEauthorrefmark{2},
        Xia Li\IEEEauthorrefmark{4}\textsuperscript{2}, 
        Xiuding Cai\IEEEauthorrefmark{1}\IEEEauthorrefmark{2}\textsuperscript{3}, 
        Xueyao Wang\IEEEauthorrefmark{1}\IEEEauthorrefmark{2}\textsuperscript{4}, 
        Yali Shen\IEEEauthorrefmark{3}\textsuperscript{5}, 
        Yu Yao\IEEEauthorrefmark{1}\IEEEauthorrefmark{2}\textsuperscript{6}
    }
    \IEEEauthorblockA{
        \IEEEauthorrefmark{1}Chengdu Institute of Computer Application, Chinese Academy of Sciences, Chengdu, China\\
        \IEEEauthorrefmark{2}School of Computer Science and Technology, University of Chinese Academy of Sciences, Beijing, China\\
        \textsuperscript{1}fulinjie19@mails.ucas.ac.cn, 
        \textsuperscript{3}caidiuding20@mails.ucas.ac.cn,
        \textsuperscript{4}wangxueyao221@mails.ucas.ac.cn, 
        \textsuperscript{6}casitmed2022@163.com\\
        \IEEEauthorrefmark{3}Sichuan University West China Hospital Department of Abdominal Oncology, China\\
        \textsuperscript{5}sylprecious123@163.com\\
        \IEEEauthorrefmark{4}Radiophysical Technology Center, Cancer Center, West China Hospital, Sichuan University, China\\
        \textsuperscript{2}lixia\_rt\_wch@scu.edu.cn
    }
}

\maketitle

\begin{abstract}
Radiation therapy is crucial in cancer treatment. Experienced experts typically iteratively generate high-quality dose distribution maps, forming the basis for excellent radiation therapy plans. Therefore, automated prediction of dose distribution maps is significant in expediting the treatment process and providing a better starting point for developing radiation therapy plans. With the remarkable results of diffusion models in predicting high-frequency regions of dose distribution maps, dose prediction methods based on diffusion models have been extensively studied. However, existing methods mainly utilize CNNs or Transformers as denoising networks. CNNs lack the capture of global receptive fields, resulting in suboptimal prediction performance. Transformers excel in global modeling but face quadratic complexity with image size, resulting in significant computational overhead. To tackle these challenges, we introduce a novel diffusion model, MD-Dose, based on the Mamba architecture for predicting radiation therapy dose distribution in thoracic cancer patients. In the forward process, MD-Dose adds Gaussian noise to dose distribution maps to obtain pure noise images. In the backward process, MD-Dose utilizes a noise predictor based on the Mamba to predict the noise, ultimately outputting the dose distribution maps. Furthermore, We develop a Mamba encoder to extract structural information and integrate it into the noise predictor for localizing dose regions in the planning target volume (PTV) and organs at risk (OARs). Through extensive experiments on a dataset of 300 thoracic tumor patients, we showcase the superiority of MD-Dose in various metrics and time consumption.  The code is publicly available at \href{https://github.com/LinjieFu-U/mamba_dose}{https://github.com/flj19951219/mamba\_dose}.
\end{abstract}


\section{Introduction}
\begin{figure}[ht]
\includegraphics[width=1\linewidth]{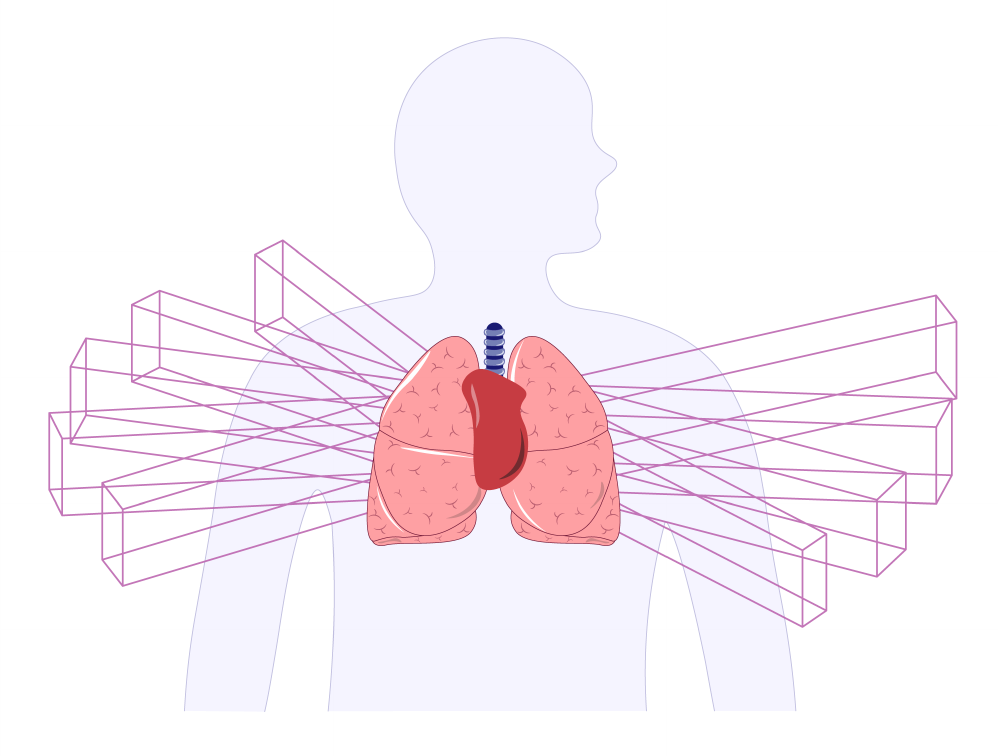}
   \caption{Demonstrate a radiation therapy plan using beam-shaped radiation.}
\label{fig1}
\end{figure}

Radiation therapy, a critical cancer treatment, necessitates precise and tailored plans to control tumors while sparing healthy tissues \cite{lee2018intensity}. Modern techniques like Intensity-Modulated Radiation Therapy (IMRT) and Volumetric Modulated Arc Therapy (VMAT) have notably enhanced treatment outcomes \cite{hussein2018automation}. They allow for precise dose sculpting by adjusting beam intensity and angles, conforming to complex tumor shapes while minimizing exposure to healthy tissues (Figure~\ref{fig1}). Nonetheless, radiotion therapy planning faces challenges: (1) anatomical changes during treatment require plan adaptation, adding complexity. (2) Collaboration between medical physicists and oncologists for plan development is time-consuming, potentially causing delays \cite{braam2006intensity}. (3) Moreover, due to individual differences and complex clinical situations, even experienced expert teams may need help to reach the optimal treatment plan quickly every time \cite{nelms2012variation}. Therefore, automated dose prediction has become particularly important. It can accelerate the treatment process, alleviate the burden on physicians, and provide a better starting point for developing treatment plans, thereby promoting more precise and effective radiation therapy.
Recent researches use deep learning to automate dose distribution map prediction. They employ complex network architectures to learn image features for this task \cite{wang2022vmat,jhanwar2022domain,jiao2023transdose,wen2023transformer,li2023multi}. However, these methods lack high-frequency detail prediction due to loss function averaging \cite{xie2023diffusion}. The diffusion model is trainable without prior data distribution knowledge and demonstrates significant potential in dose prediction \cite{feng2023diffdp,feng2023diffusion,fu2023sp}. As sampling algorithms progress, denoising network research becomes vital for diffusion models. For example, DiffDP \cite{feng2023diffdp} employs diffusion models for dose prediction, and the dose distribution maps they generate showcase enhanced high-frequency details, addressing the issue of excessive smoothing. In our previous work, we propose SP-DiffDose\cite{fu2023sp}, using a transformer-based UNet for dose prediction, outperforming  DiffDP. However, there is a heavier computational burden concerning image size due to the quadratic complexity of the self-attention mechanism in Transformers. Therefore, designing an efficient denoising network is particularly important.

Recent developments in State Space Sequence Models (SSMs) \cite{gu2023mamba,dao2024transformers}, exceptionally structured SSMs (S4) \cite{gu2021efficiently}, offer a promising solution with efficient performance in processing long sequences. The Mamba model \cite{gu2023mamba} enhances S4 through selective mechanisms and hardware optimization, performing better in dense data domains. Based on the excellent performance of Mamba in long sequence tasks, some researchers have applied Mamba to medical vision tasks, demonstrating its vast potential in modeling complex image distributions \cite{ma2024u,ruan2024vm,wang2024mamba,liu2024swin,xing2024segmamba,guo2024mambamorph}. However, research on Mamba in dose prediction is still in its early stages.

In this study, we investigate the feasibility of utilizing Mamba as a denoising network for dose prediction and propose a diffusion model called MD-Dose based on Mamba. MD-Dose consists of a forward process and a reverse process. The forward process gradually introduces noise to the original data until it becomes pure noise, while the reverse process reconstructs the dose distribution map from pure noise. To facilitate this, we develop a Mamba-structured noise predictor named Mamba-UNet to forecast the noise added at each step of the forward process, thereby generating the predicted dose map. Anatomical information provides organ structures and their relative positions. By integrating this anatomical information with noise, we assist the noise predictor in understanding dose constraints between the Planning Target Volume (PTV) and Organs at Risk (OARs), yielding more accurate dose distribution maps.

The contributions of this paper can be summarized as follows: (1) Based on the exemplary performance in the vision tasks of Mamba, we propose MD-Dose, a novel dose prediction model using Mamba as the denoising network in the diffusion model. (2) We develop a Mamba-based structural encoder to extract anatomical information from CT images and organ segmentation masks, guiding the noise predictor to generate more precise predictions. (3) MD-Dose evaluation on a clinical dataset comprising 300 patients with thoracic tumors, showing that our method achieves the best results while consuming fewer time.

\begin{figure*}[ht]
\includegraphics[width=\textwidth]{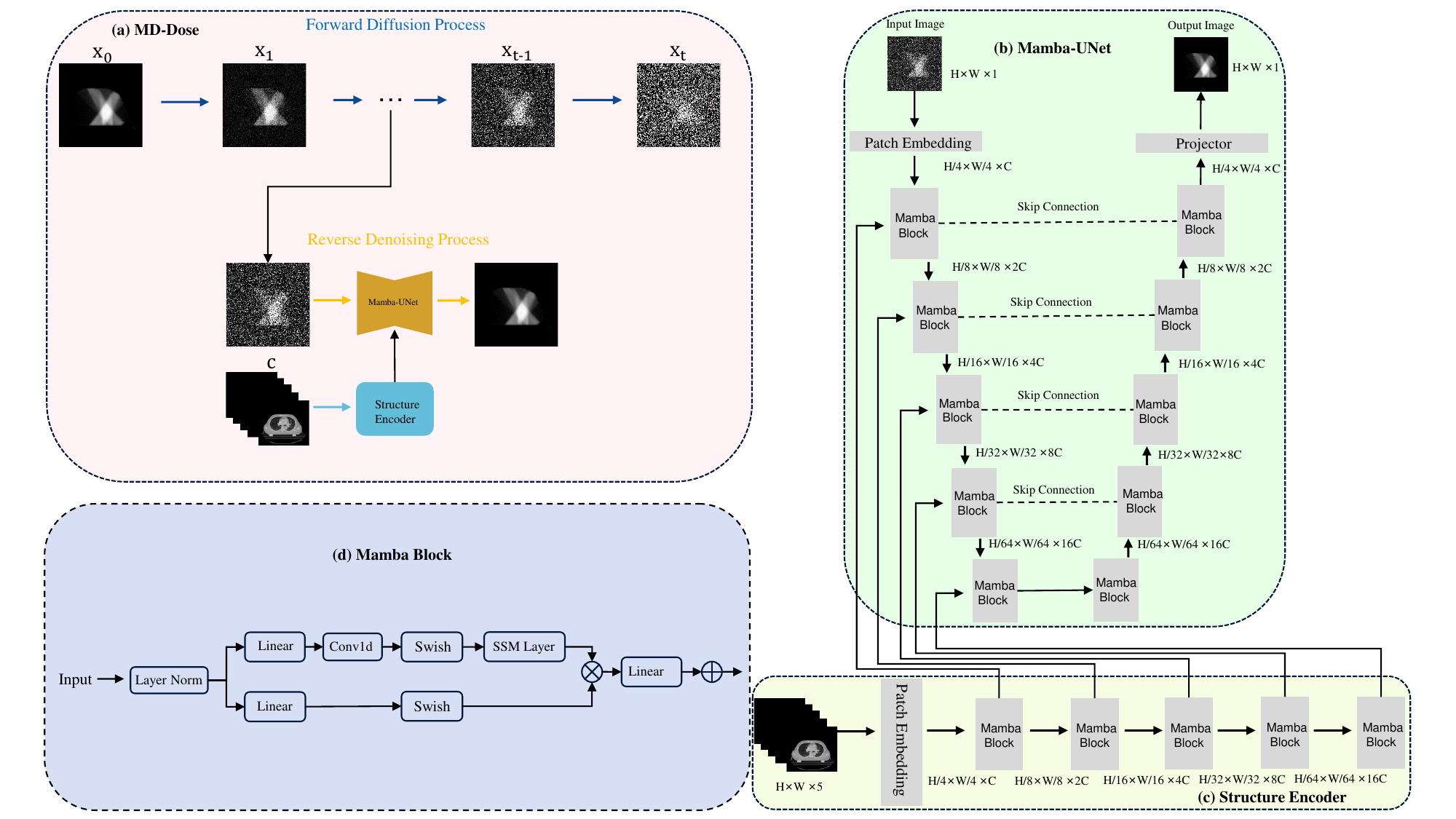}
   \caption{The overview of the proposed MD-Dose, including (a) the overall structure of MD-Dose, encompassing both the forward and backward processes of the diffusion model; (b) the proposed Mamba-UNet; (c) the proposed Structure Encoder; (d) the holistic architecture of the Mamba Block.}
\label{fig2}
\end{figure*}
\section{Methodology}
Figure \ref{fig2} presents the overall network framework of MD-Dose. (a) represents the forward noise addition and backward denoising processes of MD-Dose, (b) illustrates the network architecture of Mamba-UNet, (c) represents the structural encoder architecture, and (d) depicts the network structure of the Mamba Block. We define the dose distribution map as \(x\in \mathbb{R}^{1\times H \times W}\), structure image as \(c \in \mathbb{R}^{(2+O) \times H \times W}\), which 2 represents the image and the PTV, O represents the number of OARs, and H and W define the length and width. During the forward diffusion process, we add the Gaussian noise to the $x$ for $t$ times. In the reverse denoising process, we input $c$ to the mamba structural encoder to extract the structure feature, and fuse the structure feature with the $x_t$, finally input them into the Mamba-UNet to predict the noise in every $t$, ultimately generating accurate dose distribution maps.

\subsection{Score-based Diffusion Generative Models}
The framework of MD-Dose is designed based on Score-based diffusion generative models (SDGMs) \cite{song2020score}, which learn the distribution of data by simulating the random diffusion process of the data. MD-Dose consists of two main processes: the forward process (diffusion process) and the reverse process (denoising process).

\subsubsection{Forward Process} The forward process is a stochastic process that gradually transforms data points into random noise. The following stochastic differential equation (SDE) describes this process:
\begin{equation}
d\mathbf{x}_t = \mathbf{f}(\mathbf{x}_t, t)dt + g(t)d\mathbf{w}_t.
\end{equation}
Here, $\mathbf{x}_t$ represents the dose distribution map $x$ at time $t$, $\mathbf{f}(\mathbf{x}_t, t)$ is a drift term, $g(t)$ is the diffusion coefficient, and $\mathbf{w}_t$ is the Brownian motion.

\subsubsection{Reverse Process} The reverse process is the inverse of the forward process, aiming to reconstruct the dose distribution map $x$ from noise ${x}_t$. It can be approximated by learning a parameterized model $\mathbf{\theta}$ that attempts to reverse the diffusion process. Express the reverse process as follows:
\begin{equation}
d\mathbf{x}_t = [\mathbf{f}(\mathbf{x}_t, t, \mathbf{c}) - g^2(t, \mathbf{c})\nabla_{\mathbf{x}_t}\log p_t(\mathbf{x}_t | \mathbf{c})]dt + g(t, \mathbf{c})d\mathbf{\bar{w}}_t.
\end{equation}
Here, $p_t(\mathbf{x}_t | \mathbf{c})$ represents the probability density function of $(\mathbf{x}_t)$ at a given condition $c$, and $\mathbf{\bar{w}}_t$ corresponds to the opposite Brownian motion of $\mathbf{w}_t$, indicating the stochastic nature of the denoising process.

\subsubsection{Objective Function} During the training process, we optimize the parameters $\mathbf{\theta}$ by minimizing the reconstruction error and the negative log-likelihood of noise in the reverse process:
\begin{equation}
\begin{split}
\mathcal{L}(\theta) = \mathbb{E}_{\mathbf{x}_0, \mathbf{w}_t, \mathbf{c}}[-\log p_\theta(\mathbf{x}_0 | \mathbf{x}_t, \mathbf{c}) +\\ \lambda(t, \mathbf{c})\|\mathbf{x}_t - \mathbf{\hat{x}}_t(\mathbf{x}_0, \mathbf{w}_t, \mathbf{c}; \theta)\|^2].    
\end{split}
\end{equation}
The reconstructed data from noise $\mathbf{x}_t$ is represented by $\mathbf{\hat{x}}_t(\mathbf{x}_0, \mathbf{w}_t, c; \theta)$. $\lambda(t)$ is to balance the significance of different time steps.

\subsection{Mamba-based Denoising Network}
Inspired by the recently popular Mamba, we propose the Mamba-UNet. Mamba-UNet utilizes Mamba as a feature extraction block, adopting the encoder and decoder concept from UNet to construct a noise predictor. As shown in Figure \ref{fig2}(b), Mamba-UNet consists of three parts: 1) a Mamba encoder with multiple Mamba blocks of extracting features at different scales, 2) a Mamba decoder based on Mamba blocks for predicting the dose distribution map, and 3) skip connections link multiscale features to the decoder for feature reuse. First, we introduce the SSM layer of Mamba.

\subsubsection{SSM Layer} SSMs map the hidden state $w(t) \in \mathbb{R}^N$ to a 1-D function or sequence $y(t) \in \mathbb{R} \rightarrow x(t) \in \mathbb{R}$, which can be represented by the following linear ordinary differential equation (ODE):
\begin{equation}
\begin{aligned}
y'(t) &= Py(t) + Qw(t), \\
x(t) &= Ry(t), 
\end{aligned}
\end{equation}
where $P \in \mathbb{R}^{N \times N}$ is the state matrix, and $Q$ and $R \in \mathbb{R}^N$ are parameters, with $y'(t)  \in \mathbb{R}^N$ representing the implicit latent state.

S4 and Mamba are discrete versions of continuous systems, making them more suitable for deep learning scenarios. Specifically, S4 introduces a time scale parameter $\Delta$ and uses a fixed discretization rule to transform $A$ and $B$ into discrete parameters $\overline{P}$ and $\overline{Q}$. They are defined as follows:
\begin{equation}
\begin{aligned}
\overline{P} &= \exp(\Delta P), \\
\overline{Q} &= (\Delta P)^{-1}(\exp(\Delta P) - I) \cdot \Delta Q.  
\end{aligned}
\end{equation}
After discretizing $P$ and $Q$, linear recursion is used for rewriting:
\begin{equation}
\begin{aligned}
y_t &= \overline{P} y(t) + \overline{Q} w(t), \\
x_t &= R y(t).  
\end{aligned}
\end{equation}
Finally, the output through global convolution to calculate:
\begin{equation}
\begin{aligned}
\overline{H} &= (R\overline{Q}, C\overline{PQ}, \ldots, R\overline{P}N^{-1}\overline{Q}), \\
x &= y * \overline{H},    
\end{aligned}
\end{equation}
where $N$ is the length of the input sequence $y$, and $\overline{H} \in \mathbb{R}^M$ is a structured convolution kernel.

\subsubsection{Mamba Block} Figure \ref{fig2}(d) illustrates the comprehensive overview of the Mamba Block. Similar to the Transformer, we make the noisy image through a Patch Embedding, flattening and transposing the features with a shape of (B, C, H, W) to (B, L, C), where L = H $\times$ W, and then input it into the Mamba Block. The Mamba Block initially inputs the noisy image to a layer normalization and sends it to two parallel branches.

In the first branch, the feature is linearly expanded to (B, 2L, C) followed by successive 1D convolution layers, the Swish activation function, and the SSM layer. The second branch also expands the features to (B, 2L, C), followed by a linear layer and Swish activation function. Next, we combine the features from both branches using Hadamard multiplication. Subsequently, the features are projected back to the original shape (B, L, C), reshaped, and transposed to (B, C, H, W). Finally,  we encode the current time t and add it to the output features. Additionally, the channel count doubles after each down-sampling, while after each up-sampling, the channel count halves. 

\subsubsection{Structure Encoder} During encoding, we introduce an additional structural encoder to extract features from structural images and incorporate structural information into the noisy images to guide Mamba-UNet in restoring dose distribution maps. The structural encoder mirrors the encoder architecture of Mamba-UNet, comprising four down-sampling Mamba Blocks. As depicted in Figure \ref{fig2}(c), we feed the structural images into the structural encoder. Subsequently, we add the output of each Mamba block in the structural encoder to the corresponding Mamba block output in Mamba UNet to fuse information.

\section{Experiments}
\subsection{Datasets and Evaluation Metrics}
We conduct experiments on an in-house dataset to assess the performance of MD-Dose. The dataset includes CT images, PTV and OARs segmentation masks, and dose distribution maps from 300 patients with thoracic tumors at West China Hospital, Sichuan University, and the ethics number is ChiCTR2300074194. OARs included the heart, lungs, and spinal cord. The dataset is randomly split into training (200 patients), validation (20 patients), and test (80 patients) sets. We slice the 3D CT image into 2D slices, resize them to 256$\times$256, and utilize images with dose as inputs to the network. 

We evaluate the performance of the MD-Dose using the Dose Score \cite{liu2021cascade}, the DVH Score \cite{liu2021cascade} and the Homogeneity Index (HI) \cite{feng2023diffdp}.

The Dose Score quantifies the average relative deviation between predicted and actual dose values using the formula:
\begin{equation}
\begin{aligned}
\text{Dose Score} = \frac{1}{n} \sum_{i=1}^{n} \frac{\hat{y}_i - y_i}{y_i},   
\end{aligned}
\end{equation}
where $n$ is the sample size, $\hat{y}_i$ denotes the model's predicted dose values for the $i$-th sample, and $y_i$ represents the true dose values.

The DVH Score assesses model performance by computing the average absolute differences in DVH curves between predicted and actual doses:
\begin{equation}
\begin{split}
\text{DVH Score} = \frac{1}{n} \sum_{i=1}^{n}(\frac{1}{3}( |\hat{D_1} - D_1| + |\hat{D_{95}} - D_{95}|\\ + |\hat{D_{99}} - D_{99}|))_i, 
\end{split}
\end{equation}
where $\hat{D_{1}}$ is the minimum predicted dose value by the model, corresponding to the 1st percentile in the DVH, $\hat{D_{95}}$ is the median predicted dose value by the model, corresponding to the 95st percentile in the DVH, $\hat{D_{99}}$ is the high predicted dose value by the model, corresponding to the 99st percentile in the DVH, $D_1$ is the minimum dose value in the actual data, $D_{95}$ is the median dose value in the actual data, $D_{99}$ is the high dose value in the actual data.

The Homogeneity Index (HI) measures dose uniformity discrepancies between predicted and actual dose distributions. It quantifies uniformity by dividing the standard deviation $\sigma$ of pixel values by their mean $\mu$:
\begin{equation}
\begin{aligned}
\text{HI} = \frac{D_{2\%} - D_{98\%}}{D_{50\%}},   
\end{aligned}
\end{equation}
where $D_{2\%}$ is the dose received by 2\% of the volume, $D_{98\%}$ is the dose received by 98\% of the volume, and $D_{50\%}$ is the median dose.
\subsection{Training Details}
We implement MD-Dose using PyTorch on an NVIDIA GeForce RTX 3090. Throughout the experiment, we set the batch size to 16 and use Adam\cite{kingma2014adam} as the optimizer. The model undergoes training for 1500 epochs, with the learning rate initially set at 1e-2. It starts to decay linearly at the beginning of every epoch after 750 epochs, down to 1e-4, to accelerate convergence and prevent getting stuck in local minima. We set the parameters $\lambda_{1}$ and $\lambda_{2}$ to 1.0 and the diffusion step parameter T to 1000.
\begin{figure*}[ht!]
\includegraphics[width=\textwidth]{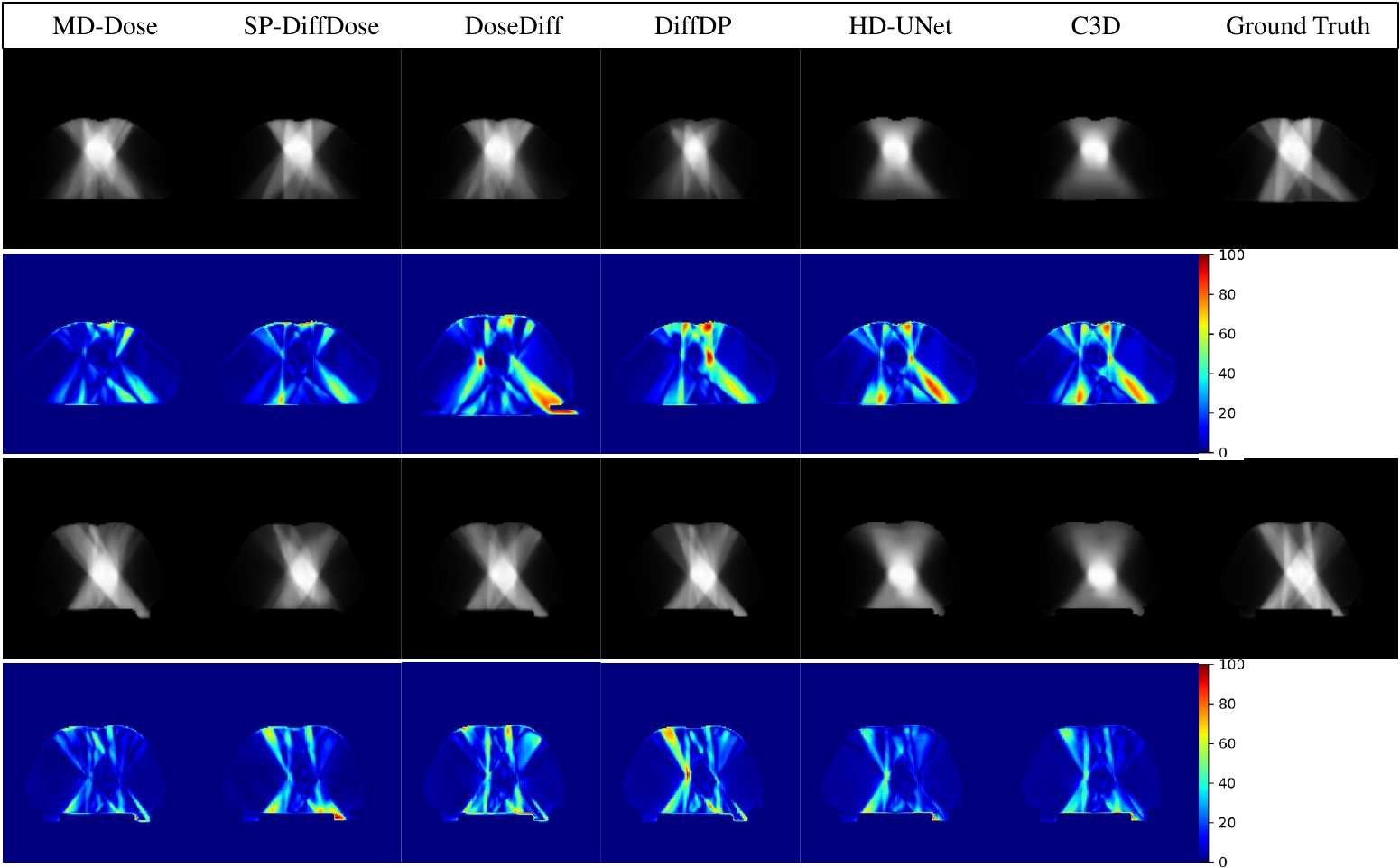}
   \caption{Visual comparisons with state-of-the-art (SOTA) methods include two sets. The first and third rows illustrate predicted dose distribution maps, and the second and fourth rows display maps depicting dose errors. The last column represents the ground truth.}
\label{fig3}
\end{figure*}

\begin{center}
\begin{table*}[ht!]
\centering
\caption{Quantitative comparison results with diffusion model methods in terms of Parameters and Inference Time. The best results are highlighted in bold. * indicates that our method significantly outperforms the compared method with a p-value of less than 0.05, as determined by a paired t-test.}
\begin{tabularx}{\textwidth}{X|XXX}
\hline
Methods & Dose Score$\downarrow$ & DVH Score$\downarrow$ & HI$\downarrow$ \\
\hline
C3D \cite{liu2021cascade} & 4.630$\pm$2.161$^*$ & 3.618$\pm$0.778$^*$ & 0.594$\pm$0.150$^*$\\
HD-UNet \cite{nguyen20193d} & 4.172$\pm$1.749$^*$ & 3.195$\pm$0.608$^*$ & 0.560$\pm$0.122$^*$\\
DiffDP \cite{feng2023diffdp}& 2.120$\pm$1.225$^*$ & 1.858$\pm$0.292$^*$ & 0.334$\pm$0.101$^*$\\
DoseDiff \cite{zhang2024dosediff}& 2.310$\pm$1.635$^*$ & 1.834$\pm$0.278$^*$ & 0.346$\pm$0.062$^*$\\
SP-DiffDose \cite{fu2023sp}& 2.000$\pm$1.131$^*$ & 1.775$\pm$0.278$^*$ & 0.292$\pm$0.068$^*$\\
MD-Dose & \textbf{1.980$\pm$1.149} & \textbf{1.572$\pm$0.239} & \textbf{0.285$\pm$0.054}\\
\hline
\end{tabularx}
\label{table1}
\end{table*}   
\end{center}
\begin{figure*}[ht!]
\includegraphics[width=\textwidth]{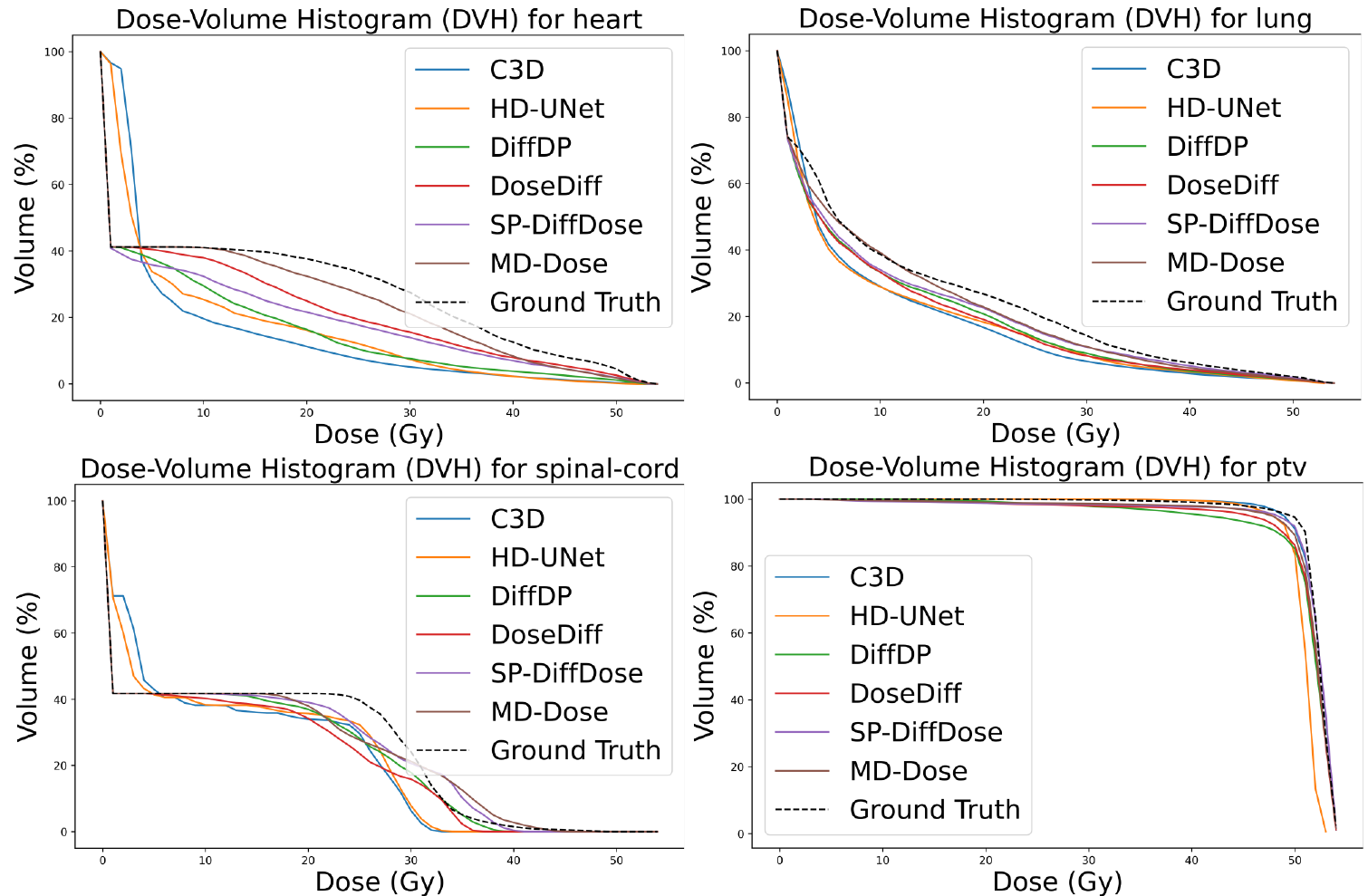}
\caption{Visualize the DVH curves of all methods, encompassing curves for the PTV, Heart, Lung, and Spinal Cord.}
\label{fig4}
\end{figure*}

\begin{table*}[ht!]
\centering
\caption{Quantitative comparison results with diffusion model methods in terms of Parameters and Inference Time. The best results are highlighted in bold.}
\begin{tabularx}{\textwidth}{X|XXXX}
\hline
 & DiffDP & DoseDiff & SP-DiffDose & MD-Dose \\
\hline
Parameter & 37.36 M & 42.88M & 84.11 M & \textbf{30.47 M}\\
Inference Time & 0.25 sec/iter & 0.27 sec/iter& 0.30 sec/iter & \textbf{0.18 sec/iter}\\
\hline
\end{tabularx}
\label{table2}
\end{table*}

\subsection{Comparison with State-Of-The-Art Methods}
To validate the effectiveness of MD-Dose, we compare it with C3D \cite{liu2021cascade}, HD-UNet \cite{nguyen20193d}, DiffDP \cite{feng2023diffdp}, SP-DiffDose \cite{fu2023sp}, and DoseDiff\cite{zhang2024dosediff}. According to the experimental results shown in Table \ref{table1}, MD-Dose surpasses the SOTA on all evaluation metrics. Specifically, compared to the C3D, MD-Dose reduces the Dose Score, DVH Score, and HI by 2.650, 2.046, and 0.309; compared to HD-UNet, these metrics decrease by 2.192, 1.623, and 0.275. These data confirm the diffusion model's advantages in dose prediction and highlight its efficiency in processing complex medical image data. Further, compared to DiffDP, MD-Dose shows reductions of 0.140, 0.286, and 0.049 in these three metrics; compared to DoseDiff, the reductions are 0.330, 0.262, and 0.061, further demonstrating MD-Dose's exceptional performance in dose prediction. Compared to SP-DiffDose, MD-Dose improves the Dose Score, DVH Score, and HI by 0.020, 0.203, and 0.007, indicating that MD-Dose exhibits superior performance over transformer-based models. Additionally, to assess whether the improvements of MD-Dose over other methods are statistically significant, this study employs a paired t-test. The experimental results in Table \ref{table3} show that the performance enhancements brought by MD-Dose are statistically substantial (p\textless0.05). These analyses confirm the superiority of MD-Dose and provide a robust scientific basis for its future clinical applications.

To thoroughly explore the predictive performance of MD-Dose, we present a series of visualization results in Figure \ref{fig3}, including the dose distribution maps predicted by various methods, the actual dose maps (GT), and the dose error maps between the predicted dose distributions and GT. The analysis reveals that the predictions from C3D and HD-UNet are overly smooth and lack high-frequency details. Although DiffDP and DoseDiff produce dose distribution maps that are closer to GT in terms of high-frequency information due to the use of convolution in their denoising networks, they primarily focus on dose prediction in tumor areas and fail to capture global information adequately, leading to less precise predictions of dose distribution in OARs. On the other hand, the Transformer-based diffusion model, SP-DiffDose, effectively captures global information, resulting in more accurate predictions of dose distribution in OARs. MD-Dose employs Mamba as its denoising network, which efficiently extracts global information and enhances computational efficiency. Figure \ref{fig3} shows that MD-Dose maintains precise high-frequency details while achieving the best visual quality. It exhibits the most minor errors in dose predictions for both tumors and OARs, demonstrating its outstanding performance in dose prediction.

The DVH curve can display the volume percentage of various dose levels within patient organs or tissues. Assessing the dose different regions receive in radiation therapy planning is crucial. Through DVH, radiation therapists can visually understand the dose distribution in each organ or tissue, helping them optimize treatment plans to ensure the best therapeutic outcomes. So, we compute DVH curves for OARs and the PTV, where closer proximity to GT indicates better prediction results. Figure \ref{fig4} illustrates that MD-Dose's DVH curves are the most comparable to GT among all OARs and the PTV.

Finally, we show the computational advantages and speed brought by Mamba. Table \ref{table2} presents the number of parameters and inference time for the four methods based on the diffusion model. Compared to DiffDP, DoseDiff, and SP-DiffDose, MD-Dose demonstrates shorter inference times with fewer parameters, indicating that MD-Dose is more efficient in dose prediction.

\begin{table*}[ht!]
\centering
\caption{The impact of denoising network selection on prediction results. Conv, Trans, and Mamba respectively represent denoising network using Convolution, Transformer, and Mamba. Mark the best results in bold.}
\begin{tabularx}{\textwidth}{XXX|XXX}
\hline
Conv & Trans & Mamba & Dose Score$\downarrow$ & DVH Score$\downarrow$ & HI$\downarrow$ \\
\hline
$\checkmark$ & & & 2.120 & 1.858 & 0.334 \\
& $\checkmark$ & & 2.000 & 1.775 & 0.292 \\
& & $\checkmark$ & \textbf{1.980} & \textbf{1.572} & \textbf{0.285} \\
\hline
\end{tabularx}
\label{table3}
\end{table*}

\begin{table*}[ht!]
\centering
\caption{The impact of structural encoder selection on prediction results. Conv-SE, Trans-SE, and Mamba-SE respectively represent structure encoder using Convolution, Transformer, and Mamba. The first row represents results without using a structural encoder. Mark the best results in bold.}
\begin{tabularx}{\textwidth}{XXX|XXX}
\hline
Conv-SE & Trans-SE & Mamba-SE & Dose Score$\downarrow$ & DVH Score$\downarrow$ & HI$\downarrow$ \\
\hline
& & & 2.076 & 1.787 & 0.298 \\
$\checkmark$ & & & 1.998 & 1.658 & 0.294 \\
& $\checkmark$ & & 1.995 & 1.627 & 0.289 \\
& & $\checkmark$ & \textbf{1.980} & \textbf{1.572} & \textbf{0.285} \\
\hline
\end{tabularx}
\label{table4}
\end{table*}

\subsection{Ablation Study}
In this section, we validate the effectiveness of Mamba and structural encoders through ablation studies. Firstly, we fix the structural encoder and experiment with different denoising networks, including Convolutional, Transformer, and Mamba. As shown in Table \ref{table3}, using Mamba as the denoising network consistently yields optimal results regardless of the structural encoder architecture, demonstrating superior efficiency. It highlights the advantage of Mamba-based denoising networks in dose prediction.
Next, we fix the denoising network as Mamba and validate the optimal selection of structural encoders. Initially, we remove the structural encoder and concatenate anatomical images with noise images as input to the denoising network. Subsequently, different backbone structure encoders extract features from anatomical structures and add them to the denoising network input and noise. As depicted in Table \ref{table4}, using a structure encoder enhances the performance across all metrics, confirming their effectiveness. Moreover, employing the Mamba architecture in the structure encoder achieves the best predictive performance, further validating the superiority of the Mamba.
\section{Discussion}
Accurate dose prediction is crucial in radiation therapy to maximize tumor control and protect OARs. However, due to the complex geometry and location of tumors, developing high-quality radiation therapy plans remains challenging. To address this issue, we propose MD-Dose, a diffusion model based on the Mamba to predict radiation dose distributions in thoracic cancer patients.

As seen in Figure \ref{fig3},  C3D and HD-UNet produce overly smooth predictions that fail to capture beam information; MD-Dose effectively captures the distribution characteristics of dose images, predicting beam directions and dose attenuation processes that align with clinical dose distributions. Meanwhile, DiffDP and DoseDiff, which use convolutional networks as denoising networks, lack the extraction of global information, resulting in less accurate dose predictions of OARs. By contrast, MD-Dose employs Mamba as its denoising network, effectively extracting global information and producing predictions more similar to actual dose distributions, particularly in OARs, as demonstrated in dose difference maps. While SP-DiffDose uses Transformers for denoising to extract global information, their attention mechanisms introduce complexity, impacting prediction efficiency. Mamba ensures that each image block only computes compressed hidden states through the corresponding scanning path, reducing complexity from quadratic to linear. As highlighted in Table \ref{table2}, this enhancement significantly improves prediction efficiency regarding model parameters and inference speed.

Optimizing radiation therapy planning involves ensuring adequate doses to tumor regions for control while minimizing damage to OARs. DVH curves intuitively display this information, aiding radiation therapists in adjusting doses and optimize plans for optimal treatment outcomes. Comparing different treatment plans or plan versions based on dose distribution differences is facilitated through DVH curves. Therefore, we demonstrate the discrepancies between DVH curves of various methods and actual DVH values. Figure \ref{fig4} shows that MD-Dose's DVH curve closely approximates actual values, showcasing its excellent capability in calculating volume percentages of different dose levels for PTV and OARs.

Future work will focus on enhancing MD-Dose's capabilities and applicability in clinical settings. It includes optimizing the Mamba architecture further to improve its performance in predicting radiation therapy dose distributions for thoracic cancer patients. Expanding experimental datasets beyond thoracic tumor patients will provide insights into the model's generalization across different cancer types and anatomical regions. This broader testing scope will validate MD-Dose's effectiveness across diverse clinical scenarios and further establish its superiority in performance metrics and computational efficiency.
\section{Conclusion}
In this paper, we propose a novel radiation dose prediction method called MD-Dose. MD-Dose utilizes Mamba as a denoising network to predict dose distribution maps for cancer patients. It also incorporates a Mamba encoder to extract structural information from anatomical images and integrate it into the denoising network, resulting in higher-quality dose distribution maps. MD-Dose can provide dose distribution maps with more high-frequency details compared to other methods, surpassing other diffusion model methods regarding inference speed. Through our approach, we can utilize the generated dose distribution maps as the initial solution for clinical radiotherapy planning, easing the burden on physicists and physicians and assisting cancer patients in undergoing more effective and precise treatment planning.

\section*{Acknowledgement}
This work was supported by Department of Science and Technology of Sichuan Province (RZHZ2022008) and 1.3.5 project for disciplines of excellence, West China Hospital, Sichuan University (20HXJS040).

\bibliographystyle{IEEEtran}
\bibliography{refer}

\vspace{12pt}

\end{document}